\documentclass[12pt]{article} 
\usepackage{sigsam-edit, amsmath}
\usepackage{graphicx}

\issue{}
\articlehead{}
\titlehead{Branch Cuts in {\sc Maple} 17}
\authorhead{England et al.}
\begin{document}

\title{Branch Cuts in {\sc Maple} 17}

\author{M.~England$^1$, E.~Cheb-Terrab$^2$, R.~Bradford$^1$, J.H.~Davenport$^1$ and D.~Wilson$^1$ \\
$^1$Department of Computer Science, University of Bath,
Bath, UK, BA2 7AY \\
$^2$Maplesoft, 615 Kumpf Drive, Waterloo, ON, Canada, N2V 1K8 \\
\texttt{ \{M.England, R.J.Bradford, J.H.Davenport, D.J.Wilson\}@bath.ac.uk} \\ \texttt{ecterrab@maplesoft.com}
}
\date{}

\maketitle

\begin{abstract}
Accurate and comprehensible knowledge about the position of branch cuts is essential for correctly working with multi-valued functions, such as the square root and logarithm.  We discuss the new tools in {\sc Maple} 17 for calculating and visualising the branch cuts of such functions, and others built up from them.  The cuts are described in an intuitive and accurate form, offering substantial improvement on the descriptions previously available.  
\end{abstract}

\noindent This work was supported by EPSRC grant EP/J003247/1.  

\section{Introduction}
\label{sec:intro}

When defining multi-valued functions (such as the natural logarithm and the square root) choices must be made for the positioning of the branch cuts.  These usually follow \cite{AS64}, or its modern counterpart \cite{OLBC2010}.  {\sc Maple} agrees for all elementary functions except arccot, for reasons explained in \cite{CDJW00}.  
%We note that a different choice would not lead to any fewer or less complicated issues. 
There are different, often unstated, viewpoints for dealing with multi-valued functions \cite{Davenport2010}.  {\sc Maple} along with most computer algebra software (and indeed most users) works with multi-valued functions by defining their single-valued counterparts which will have discontinuities over the branch cuts.  It is important that users of these functions understand the position of such cuts.

For this purpose we built the \texttt{BranchCuts} package, now integrated into {\sc Maple} 17 as part of the \texttt{FunctionAdvisor} project \cite{ChebTerrab2002}.  The code calculates and visualises the branch cuts of functions of a single complex variable.  It is used by default in {\sc Maple} 17 while readers with older versions can download a version with the basic functionality from \verb+http://opus.bath.ac.uk/32511/+.

In \cite{EBDW13} we presented in detail the generic algorithms which can be used for calculating branch cuts.  Here we focus on the practical improvements offered by the {\sc Maple} implementation.  

\section{Branch cuts in {\sc Maple}}
\label{sec:BC} 

Information on {\sc Maple}'s predefined functions may be accessed with the \texttt{FunctionAdvisor} tool: a handbook for special functions, designed to be both human and machine readable, and to process output to fit the query \cite{ChebTerrab2002}.  We can view the defining branch cuts for a given function by querying the name without argument.  For example,
\begin{verbatim}
> FunctionAdvisor(branch_cuts, ln);
\end{verbatim}
\begin{equation}
[\ln(z), z < 0]
\end{equation}
\noindent indicates the logarithm has a branch cut along the negative real axis.  

\begin{figure}[ht] 
\begin{center}
\includegraphics[width=4.7cm]{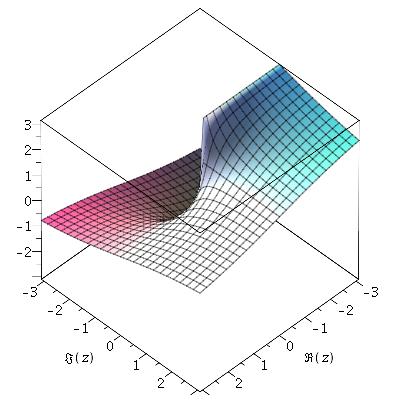}
\hspace*{0.3cm}
\includegraphics[width=4.7cm]{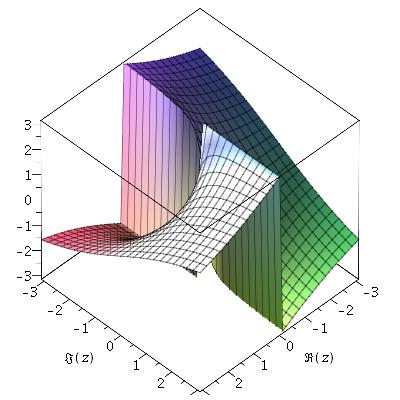}
\hspace*{0.3cm}
\includegraphics[width=4.7cm]{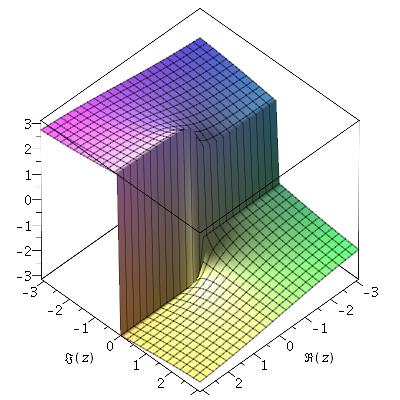}
\end{center}
\caption{Plots of the imaginary parts of the three functions discussed in Section \ref{sec:BC}.} 
\label{fig:BC}
\end{figure}

In {\sc Maple} 16 executing the command on a function with argument would lead to the same output with $z$ replaced by this argument (modulo automatic simplification).  For example,
\begin{verbatim}
> FunctionAdvisor(branch_cuts, ln(z^2));
\end{verbatim}
\begin{equation} \label{ex2}
[\ln(z^2), z^2  < 0]
\end{equation}
indicating that the function has a branch cut when $z^2$ takes negative real values.  
From this we can infer that the cuts are along the imaginary axis excluding zero.  
The statement is accurate but it would be more useful to return the imaginary axis directly.  Further, consider the example,
\begin{verbatim}
> FunctionAdvisor(branch_cuts, ln(-sqrt(z)));
\end{verbatim}
\begin{equation} \label{ex3}
[\ln(-\sqrt{z}), -\sqrt{z} < 0]
\end{equation}
from which we infer that the function has a branch cut along the positive real axis.  However, the function is also discontinuous along the negative real axis due to the extra branch cut introduced by the square root. %and so the output was incomplete.
%Numerical 3d plots of the imaginary part of these three functions are given in Figure \ref{fig:Intro}, making the discontinuities clear.  
Figure \ref{fig:BC} presents 3d plots of the imaginary part of these three functions making the discontinuities clear.  We display the imaginary part as crossing the branch cut causes the logarithm to change by $2\pi$i.  For other functions such as arctan the discontinuity occurs in the real part.  Pairs of 3d plots for the real and imaginary parts are now available directly from the \texttt{FunctionAdvisor} using the extra argument \texttt{plot=3d}.
The position of branch cuts may be inferred from such plots, but {\sc Maple} 17 is able to give accurate and intuitive algebraic descriptions and hence 2d visualisations.

\section{Improved representing and calculation of branch cuts}
\label{sec:cuts}

Two approaches have been implemented for calculating branch cuts.  The first moves the problem from one complex variable to two real variables: $z=\Re(z)+i\Im(z)$.  The resulting semi-algebraic system is then solved, usually using cylindrical algebraic decomposition (CAD), so that each branch cut is represented by an equation setting one of the variables in terms of the other and inequations bounding the independent variable.  
%For example the branch cut for the natural logarithm is given by the equation $\Im(z)=0$ and inequality $\Re(z)<0$.  
Hence in {\sc Maple} 17 the second example from Section \ref{sec:BC} becomes
\begin{verbatim}
> FunctionAdvisor(branch_cuts, ln(z^2));
\end{verbatim}
\begin{equation} 
\tag{$2'$} 
%[\ln(z^2), And(\Re(z) = 0, 0 < \Im(z)), And(\Re(z) = 0, \Im(z) < 0)]
 [\ln(z^2), \Re(z) = 0 \mbox{ And } 0 < \Im(z), \Re(z) = 0 \mbox{ And } \Im(z) < 0]
\end{equation}
\noindent which is as accurate as (\ref{ex2}) but is more useful for the user and subsequent computations.

When the argument is not polynomial a second approach is preferred where cuts are stored as a (possibly complex and transcendental) function of a real parameter with a given range, as in \cite{DF94}. 
%a contains fractional powers this approach requires that a polynomial system be obtained using a de-nesting procedure.
%, for which there is currently only a limited implementation.  
%For such examples (and also those whose branch cuts are not semi-algebraic)  a 
%So for example the branch cut for the natural logarithm is here given by $z=a$ where $a \in (-\infty,0)$.  
This is achieved by setting the argument to the parameter and inverting to find solutions for the variable using {\sc Maple}'s \texttt{solve} capabilities, applying with a range determined by the defining cut.  
We can compute the branch cuts of algebraic expressions using all the mathematical functions of the language for which the branch cuts are known.  For situations like sums and products both approaches take the union of the individual cuts.  Hence in {\sc Maple 17} the following output is obtained for the third example, which is both more intuitive and more accurate than (\ref{ex3}).
\begin{verbatim}
> FunctionAdvisor(branch_cuts, ln(-sqrt(z)));
\end{verbatim}
\begin{equation}
\tag{$3'$}
%[\ln(-\sqrt{z}), And( z = \alpha^2, \alpha \in RealRange(-\infty, 0)), z < 0]
 [\ln(-\sqrt{z}), z = \alpha^2 \mbox{ And } \alpha \in RealRange(-\infty, 0), z < 0]
\end{equation}
More details on the algorithms are available in \cite{EBDW13}.  
In {\sc Maple} 17 2d visualisations of the algebraic descriptions of the branch cuts may be obtained directly from the \texttt{FunctionAdvisor} with the optional argument \texttt{plot=2d}.  An example is on the left of Figure \ref{fig:arcsin}.  

The code considers any univariate function with defining cuts known to the \texttt{FunctionAdvisor} including elementary functions and their inverses.
%, inverse hyperbolic functions and integral functions.  
It also covers multivariate functions with branch cuts in one variable only, and hence those which come with parameters such as the Bessel functions and Chebyshev polynomials.
% and Jacobi $\theta$-functions.
% are covered, important given the increased use of such special functions in modelling.
We note that understanding the position of branch cuts can also have applications in computer algebra itself, for example in the safe application of identities.  Consider 
\begin{equation}
\label{eq:eg}
2\arcsin(z) = \arcsin(2z\sqrt{1-z^2}).
\end{equation}
A visualisation of the branch cuts of (\ref{eq:eg}) produced from the new algebraic description is given on the left of Figure \ref{fig:arcsin} with the other images plots of the imaginary part of LHS(\ref{eq:eg})$-$RHS(\ref{eq:eg}).  Identity (\ref{eq:eg}) is true within the hour glass shaped region bounded by the cuts 
%hyperbola $y^2-x^2+1=0$ 
and false otherwise.  Algorithmic approaches to identifying such regions of truth using branch cuts and CAD have been studied in a series of papers starting with \cite{BD02}.  Recent progress was reported in \cite{DBEW12} and we note that the new CAD algorithm in \cite{BDEMW13} is very well suited for dealing with input from the first branch cut representation.

\section{When is a branch cut not a branch cut?}
\label{sec:problems}

\begin{figure}[t] 
\begin{center}
\includegraphics[width=4.7cm]{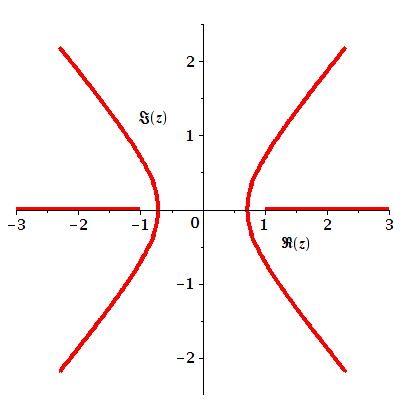}
\hspace*{0.3cm}
\includegraphics[width=4.7cm]{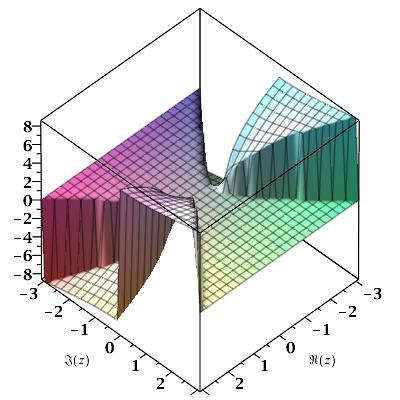}
\hspace*{0.3cm}
\includegraphics[width=4.7cm]{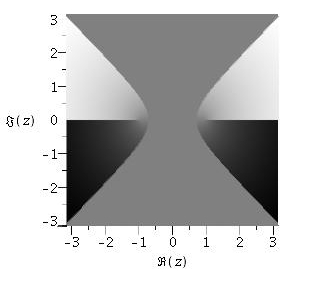}
\end{center}
\caption{Visualising the branch cuts of equation (\ref{eq:eg}).} 
\label{fig:arcsin}
\end{figure}

Failure could occur due to time/memory constraints, or the \texttt{solve} tools being unable to identify all solutions.  However, for all practical examples encountered these problems do not occur. 
A more prevalent issue is branch cuts being returned which do not correspond to discontinuities of the input.  We call these \emph{spurious cuts} and they have two causes.  First, when functions are combined the discontinuities introduced may cancel.  Second, when the argument of a function contains fractional powers then the solve tools may find solutions corresponding to all possible roots rather than the specific root concerned.  
%More details are given in \cite{EBDW13} 
In {\sc Maple} 17 a cautious approach is used, including branch cuts that may be spurious rather than risk losing true cuts.  
%This cautious approach also led to a warning system, so users may apply the algorithms widely, but are informed if the output is not guaranteed.  
%As the code is more deeply integrated many of these will be removed.
We note that spurious cuts may be identified visually for example by comparing the 2d visualisation of the algebraic output with 3d plots oriented to appear 2d (as on the right of Figure \ref{fig:arcsin}).  This orientation can be obtained interactivity with the mouse or directly from the \texttt{FunctionAdvisor} using the argument \texttt{plot=32d}.  Methods to systematically identify and remove these spurious cuts from the algebraic output are currently under development.

%For example,
%\begin{verbatim}
%> FunctionAdvisor(branch_cuts, ln(exp(z)));
%Warning, branch cuts have been returned for ln(exp(z)) but we do not guarantee 
%an exhaustive list
%\end{verbatim}
%\begin{equation}
%[\ln(\exp(z)), And(z = \ln(\alpha), \alpha \in RealRange(-\infty, 0))]
%\end{equation}
%Only the cut in the principal domain has been returned,  while the function has infinitely many.  
%Such warnings will decrease with future releases of {\sc Maple} as the the code will return the full set, or the set within a specified region.

\bibliography{CAD}

\begin{thebibliography}{10}

\bibitem{AS64}
M.~Abramowitz and I.A. Stegun.
\newblock {\em Handbook of mathematical functions}.
\newblock National Bureau of Standards, 1964.

\bibitem{BD02}
R.~Bradford and J.H. Davenport.
\newblock Towards better simplification of elementary functions.
\newblock In {\em Proceedings of ISSAC '02}, pages 16--22, 2002.

\bibitem{BDEMW13}
R.~Bradford, J.H. Davenport, M.~England, S.~McCallum, and D.~Wilson.
\newblock Cylindrical algebraic decompositions for boolean combinations.
\newblock In {\em Proceedings of ISSAC '13}, pages 125--132, 2013.

\bibitem{ChebTerrab2002}
E.S. Cheb-Terrab.
\newblock The function wizard project: A computer algebra handbook of special
  functions.
\newblock In {\em Proceedings of the Maple Summer Workshop, University of
  Waterloo}, 2002.

\bibitem{CDJW00}
R.M. Corless, J.H. Davenport, D.J. Jeffrey, and S.M. Watt.
\newblock According to {A}bramowitz and {S}tegun.
\newblock {\em SIGSAM Bulletin}, 34(3):58--65, 2000.

\bibitem{Davenport2010}
J.H. Davenport.
\newblock The challenges of multivalued ``functions''.
\newblock In {\em Intelligent Computer Mathematics}, LNCS:6167, pages 1--12, Springer,
%-Verlag, 
2010.

\bibitem{DBEW12}
J.H. Davenport, R.~Bradford, M.~England, and D.~Wilson.
\newblock Program verification in the presence of complex numbers, functions
  with branch cuts etc.
\newblock In {\em Proceedings of SYNASC '12}, pages 83--88. IEEE, 2012.

\bibitem{DF94}
A.~Dingle and R.J. Fateman.
\newblock Branch cuts in computer algebra.
\newblock In {\em Proceedings of ISSAC '94}, pages 250--257, 1994.

\bibitem{EBDW13}
M.~England, R.~Bradford, J.H. Davenport, and D.~Wilson.
\newblock Understanding branch cuts of expressions.
\newblock In 
%J.~Carette, D.~Aspinall, C.~Lange, P.~Sojka, and W.~Windsteiger, editors, 
{\em Intelligent Computer Mathematics},  LNCS:7961, pages 136--151, Springer,
% Berlin Heidelberg, 
2013.

\bibitem{OLBC2010}
W.J. Olver, D.W. Lozier, R.F. Boisvert, and C.W. Clark, editors.
\newblock {\em {NIST} Handbook of Mathematical Functions}.
\newblock Cambridge University Press, 2010.  Online version at \texttt{http://dlmf.nist.gov}

\end{thebibliography}
\bibliographystyle{plain}

\end{document}